# An Integration and Assessment of Multiple Covariates of Nonstationary Storm Surge Statistical Behavior by Bayesian Model Averaging

Tony E. Wong (anthony.e.wong@colorado.edu)

Department of Computer Science, University of Colorado at Boulder

ORCID: 0000-0002-7304-3883

## Keywords

coastal flooding, natural hazards, Bayesian statistics, uncertainty, extremes

## Key points

- Bayesian model averaging is used to generate a statistical model that accounts for multiple possible storm surge covariates and nonstationary behavior
- For a site in the southeastern United States, the model marginal likelihoods are stable across covariate time series' used, but flood hazard projections are affected by model choice
- A nonstationary model that includes four covariates raises the 100-year storm surge return level by up to 23 centimeters relative to a stationary model

## Abstract

Projections of coastal storm surge return levels are a basic requirement for effective management of coastal risks. A common approach to estimate hazards posed by extreme sea levels is to use a statistical model, which may use a time series of a climate variable as a covariate to modulate the statistical model and account for potentially nonstationary storm surge behavior (e.g., North Atlantic Oscillation index). Previous work using nonstationary statistical approaches, however, has demonstrated the importance of accounting for the many inherent modeling uncertainties. Additionally, previous assessments of coastal flood hazard using statistical modeling have typically relied on a single climate covariate, which likely leaves out important processes and leads to potential biases in the projected flood hazards. Here, I employ upon a recently developed approach to integrate stationary and nonstationary statistical models, and examine the effects of choice of covariate time series on projected flood hazard. Furthermore, I expand upon this approach by developing a nonstationary storm surge statistical model that makes use of multiple covariate time series, namely, global mean temperature, sea level, North Atlantic Oscillation index and time. I show that a storm surge model that accounts for additional processes raises the projected 100-year storm surge return level by up to 23 centimeters relative to a stationary model or one that employs a single covariate time series. I find that the total marginal model likelihood associated with each set of nonstationary models given by the candidate covariates, as well as a stationary model, is about 20%. These results shed light on how best to account for potential nonstationary coastal surge behavior, and incorporate more processes into surge projections. By including a wider range of physical process information and considering nonstationary behavior, these methods will better enable modeling efforts to inform coastal risk management.

# 1 Introduction

Reliable estimates of storm surge return levels are critical for effective management of flood risks (Nicholls and Cazenave, 2010). Extreme value statistical modeling offers an avenue to estimate these return levels (Coles, 2001). In this approach, a statistical model is used to describe the distribution of extreme sea levels. Modeling uncertainties, however, include whether or not the statistical model chosen appropriately characterizes the sea levels, and whether or not the distribution changes over time - i.e., is nonstationary (Lee et al., 2017). Process-based modeling offers a mechanistically-motivated alternative to statistical modeling (e.g., Fischbach et al., 2017; Orton et al., 2016; Johnson et al., 2013), and carries its own distinct set of modeling uncertainties. Recent efforts to manage coastal flood risk have relied heavily on statistical modeling (e.g., Oddo et al., 2017; Moftakari et al., 2017; Lopeman et al., 2015; Lempert et al., 2012). The importance of statistical modeling in managing coastal risk motivates the focus of the present study on characterizing some of the relevant uncertainties in extreme value statistical modeling of flood hazards.

Common distributions for extreme value statistical modeling of coastal storm surges include generalized extreme value (GEV) models (e.g., Karamouz et al., 2017; Wong and Keller, 2017; Grinsted et al., 2013) and a hybrid Poisson process/generalized Pareto distribution (PP/GPD) models (e.g., Wong et al., 2018; Wahl et al., 2017; Hunter et al., 2017; Buchanan et al., 2017; Cid et al., 2016; Bulteau et al., 2015; Marcos et al., 2015; Arns et al., 2013; Tebaldi et al., 2012). Approaches based on the joint probability method (for example) are another alternative to analyze extreme sea levels, although the focus of the present study is restricted to extreme value distributions (McMillan et al 2011, Haigh et al 2010a, Tawn and Vassie 1989, Pugh and Vassie 1979). In the limit of large sample sizes, the GEV distribution is the limiting distribution of a sequence of block maxima of population samples (Coles, 2001). Depending on block sizes (e.g., annual or monthly), this approach places strict limitations on the available data. By contrast, the PP/GPD modeling approach can yield a richer set of data by making use of all extreme sea level events above a specified threshold (e.g. Knighton et al 2017, Arns et al 2013). Additionally, previous studies have demonstrated the difficulties in making robust modeling choices using a GEV/block maxima approach (Ceres et al., 2017; Lee et al., 2017). These relative strengths/weaknesses of the GEV versus PP/GPD approaches motivate the present study to focus on constraining uncertainties within the PP/GPD model.

To address the question of whether or not the distribution of extreme sea levels is changing over time, many previous studies have employed nonstationary statistical models for storm surge return levels. The typical approach is to fit a spatiotemporal statistical model (e.g., Menendez and Woodworth, 2010) or to allow some climate index or variable to serve as a covariate that modulates the statistical model parameters (e.g., Wong et al., 2018; Ceres et al., 2017; Lee et al., 2017; Cid et al., 2016; Grinsted et al., 2013; Haigh et al., 2010b). The present study follows and expands upon the modeling approach of Wong et al. (2018) by incorporating nonstationarity into the PP/GPD statistical model, and providing a comparison of the projected return levels and a quantification of model goodness-of-fit under varying degrees of nonstationarity.

Relatively few studies, however, have examined the use of multiple covariates or compared the use of several candidate covariates for a particular model application (Grinsted et al., 2013). The present study tackles this issue by considering several potential covariates for extreme value models that have been used previously: global mean surface temperature (Ceres et al.; 2017; Lee et al., 2017; Grinsted et al., 2013), global mean sea level (Arns et al., 2013), North Atlantic oscillation (NAO) index (Wong et al., 2018; Haigh et al., 2010b) and time (i.e., a linear change) (Grinsted et al., 2013). To avoid potential representation uncertainties as much as possible, the attention of the present study is restricted to the Sewell's Point tide gauge site in Norfolk, Virginia, USA (NOAA, 2017a), which is within the region of study of Grinsted et al. (2013).

The present study employs a Bayesian model averaging (BMA) approach to integrate and compare various modeling choices for potential climate covariates for the statistical model for extreme sea levels (Wong et al., 2018). The use of BMA permits a quantification of model likelihood associated with each of the four candidate covariates, and illuminates important areas for future modeling efforts. BMA also enables the generation of a new model that incorporates information from all of the candidate covariates and model nonstationarity structures. The main contributions of this work is to demonstrate the ability of the BMA approach to incorporate multiple covariate time series into flood hazard projections, and to examine the value of different covariate time series. The candidate covariates used here are by no means an exhaustive treatment of the problem domain, but rather serve as a proof of concept for further exploration and to provide a characterization of the structural uncertainties inherent in modeling nonstationary extreme sea levels.

To summarize, the main questions addressed by the present study are: (1) which covariates, that have been used in previous works to modulate extreme value statistical models for storm surges, are favored by the BMA weighting? and (2) how do these structural uncertainties affect our projections of storm surge return levels? The remainder of this work is composed as follows. Section 2 describes the extreme value statistical model used here, the data sets and processing methods employed, the model calibration approach and the experimental design for projecting flood hazards. Section 3 presents a comparison of modeling results under the assumptions of the above four candidate covariates, as well as when all four are integrated using BMA. Section 4 interprets the results and discusses the implications for future study, and Section 5 provides a concluding summary of the present findings.

## 2 Methods

### 2.1 Extreme value model

First, to detrend the raw hourly sea level time series, I subtract a moving window one-year average (e.g., Wahl et al., 2017; Arns et al., 2013). Next, I compute the time series of detrended daily maximum sea levels. I use a PP/GPD statistical modeling approach, which requires selection of a threshold, above which all data are considered as part of an extreme sea level event. In an effort to maintain independence among the final data set for analysis, these events are declustered such that only the maximal event among multiple events within a given timescale is retained in the final data set. Following many previous studies, I use a declustering timescale of three days and a threshold matching the 99th percentile of the time series of detrended daily maximum sea levels (e.g., Wahl et al., 2017). The interested reader is directed to Wong et al. (2018) for further details on these methods, and to Wong et al. (2018), Wahl et al. (2017) and Arns et al. (2013) for deeper discussion of the associated modeling uncertainties.

The probability density function (pdf, $f$) for the GPD is given by:

$$f(x(t) \mid \mu(t), \sigma(t), \xi(t)) = \frac{1}{\sigma(t)}\left(1 + \xi(t)\frac{x(t)-\mu(t)}{\sigma(t)}\right)^{-(1+1/\xi(t))}$$

where $\mu(t)$ is the threshold for the GPD model (which does not depend on time t here), $\sigma(t)$ is the GPD scale parameter (meters), $\xi(t)$ is the GPD shape parameter (unitless) and $x(t)$ is sea level height at time t (processed as described above). Note that $f$ only has support for $x(t) \geq \mu(t)$, i.e., for exceedances of the threshold $\mu$. A Poisson process is assumed to govern the frequency of threshold exceedances:

$$g(n(t) \mid \lambda(t)) = \frac{(\lambda(t)\Delta t)^{n(t)}}{n(t)!} exp(-\lambda(t)\Delta t)$$

where $n(t)$ is the number of exceedances in time interval $t$ to $t+\Delta t$ and $\lambda(t)$ is the Poisson rate parameter (exceedances per day).

Following previous work, nonstationarity is incorporated into the PP/GPD parameters as:

$$\lambda(t) = \lambda_0 + \lambda_1 \varphi(t)$$

$$\sigma(t) = exp(\sigma_0 + \sigma_1 \varphi(t))$$

$$\xi(t) = \xi_0 + \xi_1 \varphi(t)$$

where $\lambda_0$, $\lambda_1$, $\sigma_0$, $\sigma_1$, $\xi_0$ and $\xi_1$ are all unknown constant parameters and $\varphi(t)$ is a time series covariate that modulates the behavior of the storm surge PP/GPD distribution (Wong et al., 2018; Grinsted et al., 2013). As in these previous works, I assume the parameters are stationary within a calendar year.

Assuming each element of the processed data set $x$ is independent and given a full set of model parameters $\theta = (\lambda_0, \lambda_1, \sigma_0, \sigma_1, \xi_0, \xi_1)$, the joint likelihood function is

$$L(x \mid \theta) = \prod_{i=1}^{N} g(n(y_i) \mid \lambda_0, \lambda_1) \prod_{j=1}^{n(y_i)} f(x_j(y_i) \mid \sigma_0, \sigma_1, \xi_0, \xi_1)$$

where $y_i$ denotes the year indexed by $i$, $x_j(y_i)$ is the $j$th threshold exceedance in year $y_i$ and $n(y_i)$ is the total number of exceedances in year $y_i$. The product over exceedances in year $y_i$ in this equation is replaced by one for any year with no exceedances.

Also following Wong et al. (2018), I consider these four potential model structures for each of four candidate covariates $\varphi(t)$: time, sea level, temperature and NAO index (discussed in greater detail below). Note that if $\lambda_1 = \sigma_1 = \xi_1 = 0$, the PP/GPD parameters $\lambda(t)$, $\sigma(t)$ and $\xi(t)$ are constant, yielding a stationary statistical model. This model is denoted "ST." If the frequency of threshold exceedances is permitted to be nonstationary, then $\sigma_1 = \xi_1 = 0$, but $\lambda_1$ is not necessarily equal to zero. This model permits one parameter, $\lambda$, to be nonstationary, and is denoted "NS1." Similar models are constructed by permitting both λ and σ to be nonstationary while holding $\xi_1 = 0$ (NS2) and permitting all three parameters to be nonstationary (NS3).

A set of 13 total candidate models is generated by considering each of these four candidate model structures for each of the four candidate covariates; model ST is the same for all covariates. For each of the 13 candidate models, I use ensembles of PP/GPD parameters, calibrated using observational data and forced using time series for the appropriate covariate, to estimate the 1:100 storm surge return level for Norfolk in 2065 (the surge height corresponding to a 100-year return period). Projections for other return periods are available in the Supplemental Material accompanying this article.

## 2.2 Data

The tide gauge station selected for this study is Sewells Point (Norfolk), Virginia, United States (NOAA, 2017a). Norfolk was selected for two reasons. First, the Norfolk tide gauge record is long and nearly continuous (89 years). Second, Norfolk is within the southeast region of the United State considered by Grinsted et al. (2013), so the application of global mean surface temperature as a covariate for changes in storm surge statistical characterization is reasonable. This assumption should be examined more closely if these results are to be interpreted outside this region. It is important to make clear that the assumption of a model structure in which storm surge parameters covary with some time series $\varphi$ does *not* imply the assumption of any direct causal relationship. Rather, the use of a covariate $\varphi$ to modulate the storm surge is meant to take advantage of dependence relationships among the covariate time series and storm surge. For example, an unknown mechanism could lead to changes in both global mean temperature as well as storm surge return levels. The fact that temperature does not directly cause the change in storm surge does *not* mean that temperature is not a useful indicator of changes in storm surge. That is why this work has chosen the term "covariate" for these time series.

The time covariate is simply the identity function. For example, $\varphi(1928) = 1928$ for the year $y_1=1928$. The nonstationary model assuming a time covariate corresponds to the linear trend model considered by Grinsted et al. (2013).

For the NAO index covariate time series, I use as historical data the monthly NAO index data from Jones et al. (1997), and as projections the MPI-ECHAM5 sea level pressure projection under SRES scenario A1B as part of the ENSEMBLES project (www.ensembles-eu.org; Roeckner et al., 2003). As forcing to the nonstationary models, I calculate the winter mean (DJF) NAO index following Stephenson et al. (2006).

For the temperature time series, I use as historical data the annual global mean surface temperatures from the National Centers for Environmental Information data portal (NOAA, 2017b), and as projections the CNRM-CM5 simulation (member 1) under Representative Concentration Pathway 8.5 (RCP8.5) as part of the CMIP5 multi-model ensemble (http://cmip-pcmdi.llnl.gov/cmip5/).

For the sea level time series, I use as historical data the global mean sea level data set of Church and White (2011). For projecting future flood hazard, I use the simulation from Wong and Keller (2017) yielding the ensemble median global mean sea level in 2100 under RCP8.5.

Each of the covariate data records and the tide gauge calibration data record are trimmed to 1928-2013 (86 years) because this is the time period for which observational data for all time series are available. I normalize all of the covariate time series so that the minimum/maximum range for the historical period is 0 to 1; the projections period (to 2065) may lie outside of the 0-1 range. Thus, all candidate models are calibrated to the same set of observational data, and the covariate time series are all on the same scale, making for a cleaner comparison.

## 2.3 Model calibration

I calibrate the model parameters using a Bayesian parameter calibration approach (e.g., Higdon et al., 2004). As prior information $p(\theta)$ for the model parameters, I select 27 tide gauge sites with at least 90 years of data available from the University of Hawaii Sea Level Center data portal (Caldwell et al., 2015). I process each of these 27 tide gauge data sets and the Norfolk data that is the focus of this study as described in Section 2.1. Then, I fit maximum likelihood parameter estimates for each of the 13 candidate model structures. For each model structure and for each parameter, I fit either a normal or gamma prior distribution to the set of 28 maximum likelihood parameter estimates, based on whether the parameter support is infinite (in the case of $\lambda_1$, $\sigma_1$, $\xi_0$ and $\xi_1$) or half-infinite (in the case of $\lambda_0$ and $\sigma_0$).

The essence of a Bayesian calibration approach is to use Bayes' theorem to combine the prior information $p(\theta)$ with the likelihood function $L(x \mid \theta)$ as the posterior distribution of the model parameters $\theta$, given the data $x$:

$$p(\theta \mid x) \propto L(x \mid \theta) \, p(\theta).$$

I use a robust adaptive Metropolis-Hastings algorithm to generate Markov chains whose stationary distribution is this posterior distribution (Vihola, 2012), for each of the 13 distinct model structures (level of nonstationarity/parameter covariate time series combinations). For each distinct model structure, I initialize each Markov chain at maximum likelihood parameter estimates, and iterate the Metropolis-Hastings algorithm 100,000 times, for ten parallel Markov chains. I use Gelman and Rubin diagnostics to assess convergence and remove a burn-in period of 10,000 iterates (Gelman and Rubin, 1992). From the remaining set of 900,000 Markov chain iterates (pooling all ten parallel chains), I draw a thinned sample of 10,000 sets of parameters for each of the distinct model structures to serve as the final ensembles for analysis.

## 2.4 Bayesian model averaging

In the context of using statistical modeling to estimate flood hazards, there has been some debate over how best to use the limited available information to constrain projections. More complex model structures can incorporate potentially nonstationary behavior (i.e., model NS1-3), but the additional parameters to estimate come at the cost of requirements of more data (Wong et al., 2018). Some work has focused on the timescale on which nonstationary behavior may be detected (Ceres et al., 2017) and others have focused on the ability of modern calibration methods to identify correct storm surge statistical model structure (Lee et al., 2017). Methods such as processing and pooling tide gauge data into a surge index permits a much richer set of data with which to constrain additional parameters (Grinsted et al., 2013) but the "best" way to reliably process data and make projections remains unclear (Lee et al., 2017). Indeed,

Lee et al. (2017) demonstrated that even the surge index methodology of Grinsted et al. (2013), which assimilates data from six tide gauge stations, likely cannot appropriately identify a fully nonstationary (NS3) model with a global mean temperature covariate. In summary, there is a large amount of model structural uncertainty surrounding model choice (Lee et al., 2017) and the model covariate time series (Grinsted et al., 2013).

Bayesian model averaging (BMA; Hoeting et al., 1999) offers an avenue to handle these concerns by combining information across candidate models, and weighting the estimates from each model by the degree to which that model is persuasive relative to the others. Using BMA, each candidate model $M_k$ is assigned a weight that is its model marginal likelihood $p(M_k \mid x)$. Each model $M_k$ yields an estimated return level in year $y_i$, $RL(y_i \mid M_k)$. The BMA estimate of the return level can then be written as an average of the return levels as estimated by each candidate model, weighted by each model's BMA weight:

$$RL(y_i \mid x) = \sum_{k=1}^{m} RL(y_i \mid M_k) \, p(M_k \mid x),$$

where $m$ is the total number of models under consideration. The BMA weights for each model $M_k$ are given by Bayes' theorem and the Law of Total Probability as:

$$p(M_k \mid x) = \frac{p(x \mid M_k) \, p(M_k)}{\sum_{j=1}^{m} p(x \mid M_j) \, p(M_j)}.$$

The prior distribution over the candidate models is assumed to be uniform ($p(M_i) = p(M_j)$ for all $i, j$). The probabilities $p(x \mid M_k)$ are estimated using bridge sampling and the posterior ensembles from the Markov chain Monte Carlo analysis (Meng and Wing, 1996).

For each of the four covariate time series, in addition to the ensembles of 100-year storm surge returns levels for each of the four candidate models, I produce a BMA ensemble of 100-year return levels as outlined above. In a final experiment, I pool all 13 distinct model structures to create a BMA ensemble in consideration of all levels of nonstationarity and covariate time series. This BMA-weighted ensemble constitutes a new model structure that takes into account more mechanisms for modulating storm surge behavior - time, temperature, sea level and NAO index. This experiment has two aims: (1) to assess the degree to which the Norfolk data set informs our choice of covariate time series and (2) to quantify the impacts of single-model or single-covariate choice in the projection of flood hazards.

## 3 Results

### 3.1 BMA weights for individual models

The BMA weights for the stationary model (ST) and each of the three nonstationary models (NS1-3) are robust across changes in the covariate time series employed to modulate the storm surge model parameters (Figure 1). The ST model receives about 55% weight, the NS1 model (where the Poisson rate parameter $\lambda$ is nonstationary) receives about 25% weight, the NS2 model (where both $\lambda$ and $\sigma$ are nonstationary) receives about 15% weight, and the fully nonstationary NS3 model receives about 5% weight. While the stationary model is consistently has the highest model marginal likelihood, the fact that the nonstationary models have appreciable weight associated with them is a clear signal that these processes should not be ignored. In light of these results, it becomes rather unclear which is the "correct" model choice, and which covariate is the most appropriate. The latter question will be addressed in Sections 3.3 and 3.4. The former question is addressed using BMA to combine the information across all of the candidate model structures, for each covariate individually. In this way, BMA permits the use of model structures which may have a large uncertainties but are still useful to inform risk management strategies.

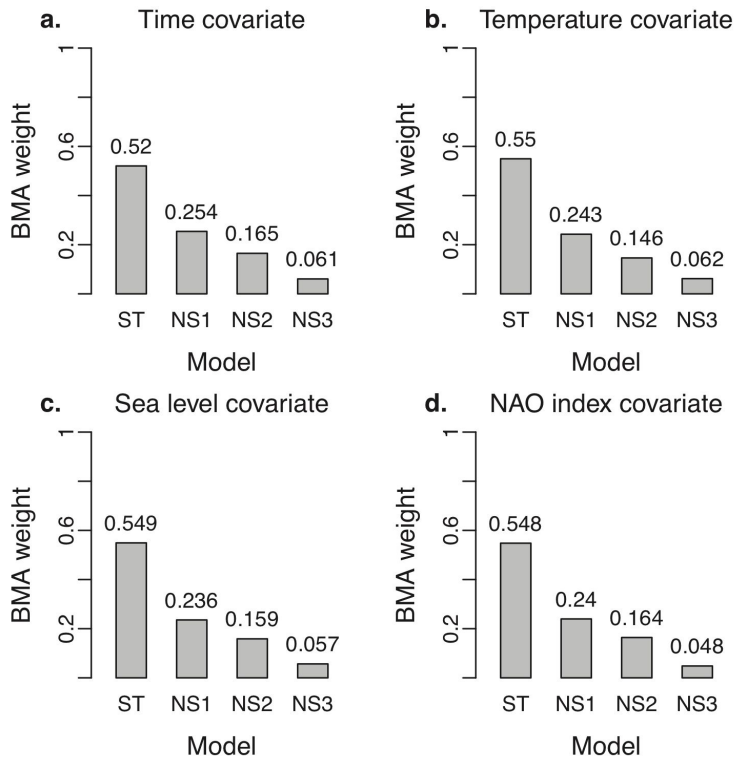

**Figure 1.** Bar plots showing the Bayesian model averaging weight for each of the four candidate models (ST, NS1, NS2 and NS3) using as a covariate: (a) time, (b) temperature, (c) sea level and (d) NAO index.

### 3.2 Return levels for individual models

When BMA is used to combine all four candidate ST/nonstationary models for each candidate covariate, the ensemble median projected 100-year return level in 2065 increases by between 4 and 23 centimeters, depending on the covariate used (Figure 2). Interestingly, the use of BMA with a global mean temperature or sea level covariate widens the uncertainty range relative to the stationary model (Figure 2b, c), whereas the BMA-weighted ensembles using time or NAO index as a covariate tightens the uncertainty range. By considering nonstationarity in the PP/GPD shape parameter, model NS3 consistently displays the widest uncertainty range for 100-year return level, and a lower posterior median than a stationary model. This indicates the large uncertainty associated with the GPD shape parameter.

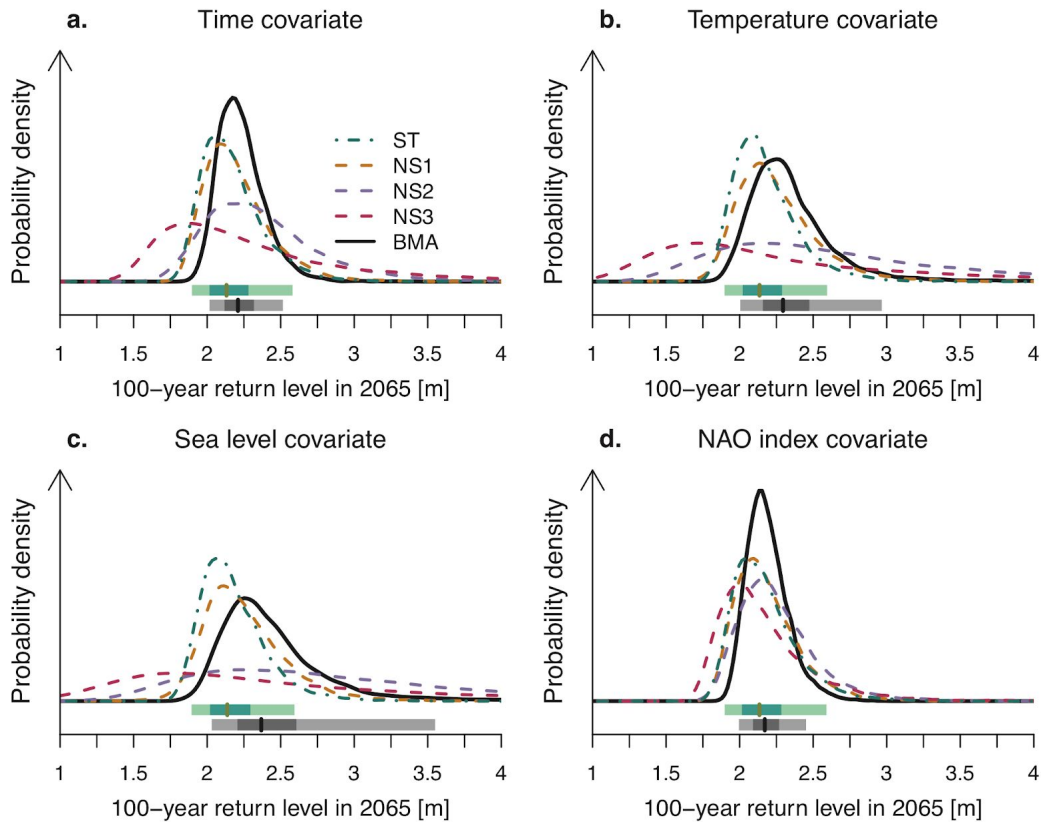

**Figure 2.** Empirical probability density functions for the 100-year storm surge return level level (meters) at Norfolk, Virginia, as estimated using one of the four candidate model structures and using the Bayesian model averaging ensemble. Shown are nonstationary models where the statistical model parameters covary with (a) time, (b) global mean surface temperature, (c) global mean sea level and (d) winter mean NAO index. The bar plots provide the 90% credible range (lightest shading), the interquartile range (moderate shading) and the ensemble medians (dark vertical line).

### 3.3 BMA weights all together

When all 13 distinct model structures are considered in the BMA weighting simultaneously, the models' BMA weights display a clear trend in favor of less complex structures (Figure 3). If one wishes to use these results to select a single model for projecting storm surge hazard, then, based on BMA weights, a stationary model would be the appropriate choice. In light of the results of Section 3.1, it is not surprising that the fully nonstationary models (NS3) are the poorest choices as measured by BMA weight. The models are assumed to all have uniform prior probability of 1/13 (about 0.077). So, these results may be interpreted as stronger evidence for use of the stationary and NS1 models for modulating storm surges, and weaker evidence for incorporating nonstationarity in, for example, the GPD shape parameter (NS3).

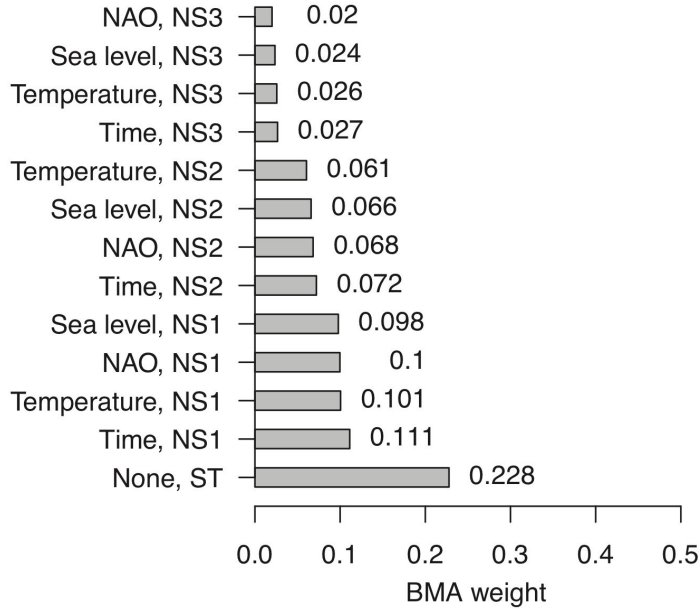

**Figure 3.** Bar plot showing the Bayesian model averaging weights for each of the 13 distinct candidate model structures, when all are simultaneously considered.

A quantification of the total marginal likelihood for each candidate covariate time series is given by adding up the BMA weights associated with each covariate's nonstationary models (NS1, NS2 and NS3) (Table 1). A stationary model has the highest total BMA weight (0.23), followed by a simple linear change in PP/GPD parameters (0.21); temperature, sea level and NAO index covariates all have roughly equal total weights (0.19). Overall, the fact that a stationary model has only 23% of the marginal model weight highlights the importance of accounting for nonstationarity.

| **Covariate:** | **Time** | **Temperature** | **Sea level** | **NAO index** | **None (ST)** |
|---|---|---|---|---|---|
| **BMA weight:** | 0.210 | 0.187 | 0.187 | 0.188 | 0.228 |

**Table 1.** Total BMA weight associated with each candidate covariate's nonstationary models and a stationary model (ST) for the full set of 13 candidate models all considered simultaneously in the BMA weighting.

### 3.4 Return levels all together

The BMA model that considers all of the covariate time series (weighted by model marginal likelihood) has a median projected 2065 storm surge return level of 2.36 m (90% credible range of 2.14-3.10 m)

(Figure 4). This more detailed model has a higher central estimate (2.36 m) of flood hazard than all of the single-covariate models' BMA ensembles except for sea level (2.16 m for NAO index, 2.21 m for time, 2.29 m for temperature, and 2.37 m for sea level). When considering the set of multi-model, single-covariate BMA ensembles (Figure 4, dashed colored lines) and the multi-model, multi-covariate BMA ensemble (Figure 4, solid black line), there is a substantial amount of uncertainty in these projections of flood hazard attributable to model structure, in particular, with regard to the choice of covariate time series.

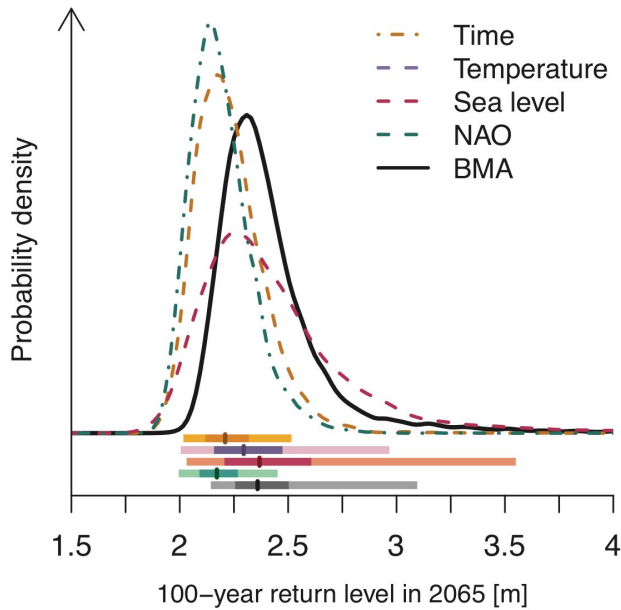

**Figure 4.** Empirical probability density functions for the 100-year storm surge return level level (meters) at Norfolk, Virginia, as estimated using the Bayesian model averaging and one of the four candidate covariates, and using the overall Bayesian model averaging ensemble with all 15 distinct candidate model structures. The bar plots provide the 90% credible range (lightest shading), the interquartile range (moderate shading) and the ensemble medians (dark vertical line). The shaded bar plots follow the same order as the legend.

## 4 Discussion

This study has presented and expanded upon an approach to integrate multiple streams of information to modulate storm surge statistical behavior (covariate time series) and to account for the fact that the "correct" model choice is almost always unknown. This approach improves on the current status of storm surge statistical modeling by accounting for more processes (multiple covariates and model structures), thereby raising the upper tail of flood hazard (by up to 23 cm for Norfolk), while constraining these additional processes using BMA. These methods will be useful, for example, in rectifying disagreement between previous assessments using nonstationary statistical models for storm surges (e.g., Lee et al.,

2017; Grinsted et al., 2013). The results presented here are consistent with those of Wong et al. (2018), who employed a single covariate BMA model based on NAO index. Both studies demonstrate that the neglect of model structural uncertainties surrounding model choices lead to an underestimation of flood hazard.

These results are in agreement with the work of Lee et al. (2017) and highlight the importance of carefully considering the degree to which more complex model structures can be relied upon against the necessity to consider more complex physical mechanisms (i.e., nonstationary storm surges). If a single-model choice is to be made, then a stationary model may be the natural choice (Table 1). However, this work provides guidance on incorporating nonstationary processes to a degree informed by the model marginal likelihoods in light of the available data. Using the full multi-model, multi-covariate BMA model substantially raises the center (from 2.13 m to 2.36 m) and upper tail (95th percentile from 2.58 m to 3.10 m) of the distribution of 100-year flood hazard relative to using a stationary model.

Of course, some caveats accompany this analysis. The covariate time series are all deterministic model input. In particular, the temperature and sea level time series do not include the sizable uncertainties in projections of these time series, which in turn depend largely on deeply uncertain future emissions pathways. The accounting and propagation of uncertainty in the covariate time series would be an interesting avenue for future study, but is beyond the scope of this work. Furthermore, this study only considers derivatives of PP/GPD model structures and does not address the deep uncertainty surrounding the choice of statistical model (Wahl et al., 2017).

This study also focuses on a single tide gauge station (Sewells Point/Norfolk, Virginia). This choice was made in light of the deep uncertainty surrounding how best to process and combine information across stations into a surge index (Lee et al., 2017) and because the Norfolk site is within the region studied by Grinsted et al. (2013), so application of global mean surface temperature as a covariate is reasonable. To extend these results to regions outside the southeastern part of the United States is an important area for future work. A key strength of the fully nonstationary multi-covariate BMA model is that the methods can be applied to any site, and the model marginal likelihoods will allow the data to inform the weight placed on the different stationary/nonstationary models and covariates.

The present study is not intended to be the final word on model selection or projecting storm surge return levels. Rather, this work is intended to present a new approach to generate a model that accounts for more processes and modeling uncertainties, and demonstrate its application to an important area for flood risk management. This study only presents a handful out of many potentially useful covariates for storm surge statistical modeling (e.g., Grinsted et al., 2013). Future work should build on the methods presented here and can incorporate other mechanisms known to be important local climate drivers for specific applications.

## 5 Conclusions

This study has presented a case study for Norfolk, Virginia, that demonstrates of the use of BMA to integrate flood hazard information across models of varying complexity (stationary versus nonstationary) and modulating model parameters using multiple covariate time series. This work finds that for the

Norfolk site, all of the candidate covariates yield similar degrees of confidence in the (non)stationary model structures, and the overall BMA model that employs all four candidate covariates projects a higher flood hazard in 2065. These results provide guidance on how best to incorporate nonstationary processes into flood hazard projections, and a framework to incorporate other locally important climate variables, to better inform coastal risk management practices.

## Acknowledgments

I gratefully acknowledge Klaus Keller, Vivek Srikrishnan and Dale Jennings for fruitful conversations. All codes are freely available from github.com/tonyewong/covariates. The author declares that there is no conflict of interest. This work was co-supported by the National Science Foundation through the Network for Sustainable Climate Risk Management (SCRiM) under NSF cooperative agreement GEO-1240507; and the Penn State Center for Climate Risk Management. Any opinions, findings, and conclusions or recommendations expressed in this material are those of the author and do not necessarily reflect the views of the National Science Foundation.

I acknowledge the World Climate Research Programme's Working Group on Coupled Modelling, which is responsible for CMIP, and I thank the climate modeling groups (listed in Table S1 of this paper) for producing and making available their model output. For CMIP the U.S. Department of Energy's Program for Climate Model Diagnosis and Intercomparison provides coordinating support and led development of software infrastructure in partnership with the Global Organization for Earth System Science Portals. The ENSEMBLES data used in this work was funded by the EU FP6 Integrated Project ENSEMBLES (Contract number 505539) whose support is gratefully acknowledged.

# Supplementary Material

| Modeling Center (or Group) | Institute ID | Model Name |
|---|---|---|
| Centre National de Recherches Météorologiques / Centre Européen de Recherche et Formation Avancée en Calcul Scientifique | CNRM-CERFACS | CNRM-CM5 |

Table S1.  CMIP5 model(s) employed in the present study.

| Return period (years) | 2.5% | 5% | 25% | 50% | 75% | 95% | 97.5% |
|---|---|---|---|---|---|---|---|
| 2 | 1.231 | 1.240 | 1.265 | 1.286 | 1.309 | 1.351 | 1.366 |
| 5 | 1.418 | 1.429 | 1.465 | 1.494 | 1.531 | 1.598 | 1.624 |
| 10 | 1.567 | 1.580 | 1.627 | 1.667 | 1.717 | 1.816 | 1.865 |
| 20 | 1.724 | 1.740 | 1.800 | 1.855 | 1.923 | 2.079 | 2.184 |
| 50 | 1.942 | 1.965 | 2.050 | 2.129 | 2.234 | 2.554 | 2.870 |
| 100 | 2.114 | 2.144 | 2.256 | 2.360 | 2.504 | 3.091 | 3.800 |
| 200 | 2.298 | 2.333 | 2.479 | 2.615 | 2.814 | 3.920 | 5.350 |
| 500 | 2.553 | 2.601 | 2.800 | 2.993 | 3.290 | 5.771 | 10.055 |
| 1000 | 2.757 | 2.819 | 3.067 | 3.317 | 3.724 | 8.259 | 17.047 |

Table S2.  Quantiles of the estimated storm surge return levels (meters) for Norfolk (Sewells Point) in 2065 using the full nonstationary multi-covariate BMA model.

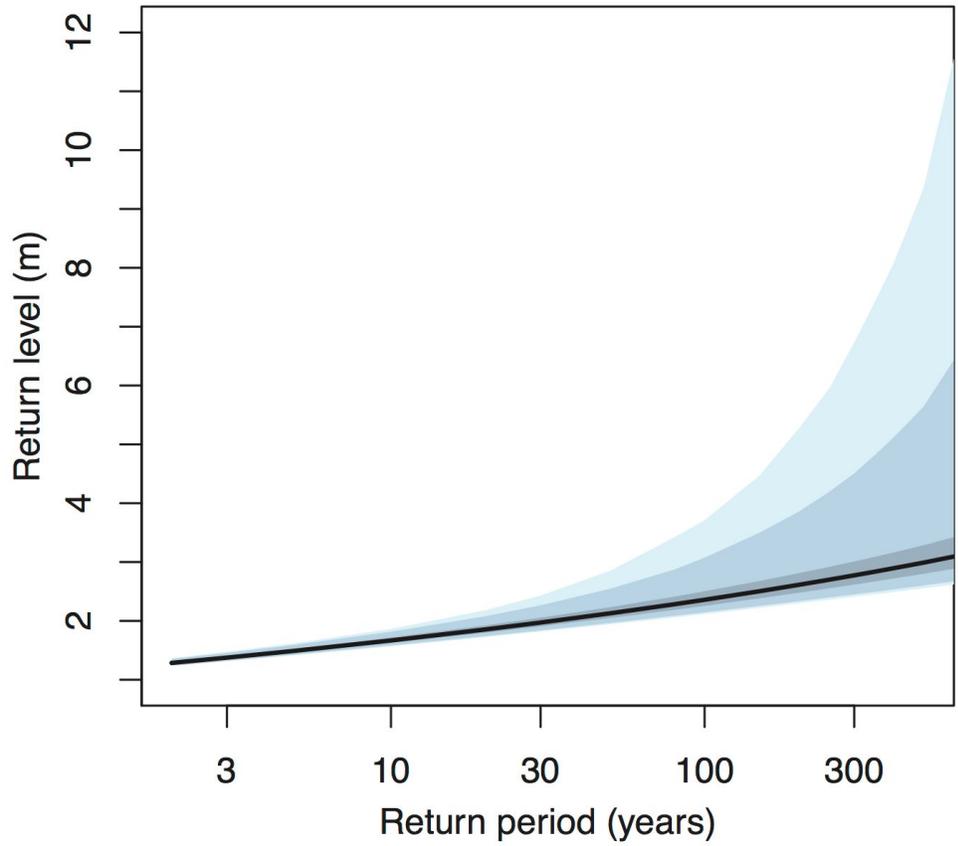

Figure S1. Storm surge return periods (years) and associated return levels (meters) in 2065 for Norfolk, using the full nonstationary multi-covariate BMA model.